\documentclass[aps,twocolumn,floatfix,footinbib]{revtex4}
\usepackage{graphicx}
\usepackage{amsmath}
\usepackage{color}
\begin{document}

\title{Nanometer-scale sharpness in corner-overgrown heterostructures}

\author{L. Steinke$^{\star}$}
\affiliation{Walter Schottky Institut, Technische Universit\"at M\"unchen, D-85748 
Garching, Germany}

\author{P. Cantwell, D. Zakharov, E. Stach}
\affiliation{School of Materials Engineering, Purdue University, West Lafayette, IN 47907, USA}

\author{N. J. Zaluzec}
\affiliation{Argonne National Laboratory,  Electron Microscopy Center, Materials Science Div., Argonne, IL 60439 USA}

\author{A. Fontcuberta i Morral, M. Bichler, G. Abstreiter}
\affiliation{\it Walter Schottky Institut, Technische Universit\"at M\"unchen, D-85748 
Garching, Germany}

\author{M. Grayson$^{\dagger}$}
\affiliation{Department of Electrical Engineering and Computer Science, Northwestern University, Evanston, IL 60208, USA and Walter Schottky Institut, Technische Universit\"at M\"unchen, D-85748 
Garching, Germany}
\date{14 July 2008}

\begin{abstract}
A corner-overgrown GaAs/AlGaAs heterostructure is investigated with transmission and scanning transmission electron microscopy, demonstrating self-limiting growth of an extremely sharp corner profile of 3.5 nm width. In the AlGaAs layers we observe self-ordered diagonal stripes, precipitating exactly at the corner, which are regions of increased Al content measured by an XEDS analysis. A quantitative model for self-limited growth is adapted to the present case of faceted MBE growth, and the corner sharpness is discussed in relation to quantum confined structures. We note that MBE corner overgrowth maintains nm-sharpness even after microns of growth, allowing the realization of corner-shaped nanostructures.
\end{abstract} 
\maketitle

The corner-overgrowth technique has been demonstrated previously \cite{one_two} to create a 90 degree junction between high-mobility
two-dimensional electron systems (2DES).  In a magnetic field, this system realizes a unique sort of
quantum Hall effect boundary state with either co- or counter-propagating edge modes at various filling factors \cite{3,4}.  In all cases, the sharpness of the confinement potential is essential to understanding the electronic structure of the device. In this letter, high- resolution transmission and scanning transmission electron microscopy (TEM/STEM) as well as simultaneous  X-ray  energy dispersive spectroscopy (XEDS) measurements on corner-overgrown heterostructures quantify the sharpness of the corner profile and give further insight into the growth mechanisms at such a corner junction.
\begin{figure}
\includegraphics{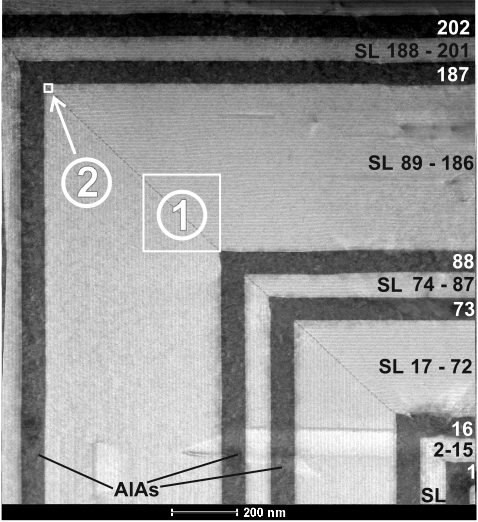}
\caption{HAADF  STEM image of the heterostructure, demonstrating sharp corners and equal layer thicknesses on both overgrown facets.  The grown layer sequence consists of 71 nm wide AlAs layers separated by GaAs/AlGaAs superlattices, where the numbers on the right count the layer index. The superlattice in region 1 is displayed at higer magnification in Fig. \ref{Fig2} a, and a high resolution bright field image of region 2 is shown in Fig. \ref{Fig3} a.}
\label{Fig1}
\end{figure}

\begin{figure}
\includegraphics{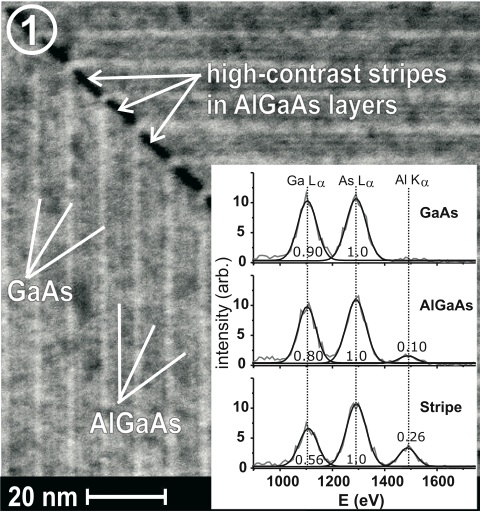}
\caption{HAADF STEM image of high contrast stripes only along the diagonal of AlGaAs layers in the superlattice (region 1 in Fig. \ref{Fig1}). The inset shows XEDS spectra from three different regions, obtained using a $\sim 1$ nm probe: GaAs, AlGaAs and one of the diagonal stripes. The numbers next to the peaks indicate the measured intensities, normalized to the As L$_{\alpha}$ intensity of each spectrum.}
\label{Fig2}
\end{figure}

A detailed description of the growth technique is found in Refs. \cite{one_two}.
The samples were grown in an ultra-high vacuum Epi GEN-II molecular beam epitaxy (MBE) system ($P < 10^{-9}$ Pa). GaAs $(110)$ substrate pieces were cleaved ex-situ to create  atomically sharp corners between two orthogonal $(110)$ 
and $(1\bar{1}0)$ facets, and mounted inside the MBE chamber with the corner facing the molecular flux. After oxide desorption the samples were overgrown under rotation ($d\alpha/dt = 5$ rpm) at a substrate temperature of $T_{sub} = 460 \, ^{\circ}$C and an As$_4$ pressure of $3.5\times10^{-3}$ Pa. Fig. \ref{Fig1} illustrates the layer sequence of the heterostructure, which consists of $71$ nm wide AlAs layers separated by $\rm Al_{0.3}Ga_{0.7}As/GaAs$ superlattices with period (8.0 nm / 2.3 nm). The AlAs layers were grown to obtain high contrast images of the corner structure in both scanning and transmission electron microscopes.

The specimen was prepared for TEM using an FEI xT Nova NanoLab Dual Beam
Focused Ion Beam / Scanning Electron Microscope (FIB/SEM). During FIB
preparation, a variation of the liftout technique \cite{5} was used to
extract an approximately 1 $\mu$m wide cross-section of the corner overgrown
heterostructure. The cross-section was then attached to a copper TEM
half-grid using platinum and further thinned to $\sim 100$ nm. Microanalytical  and high angle annular dark
field (HAADF)  STEM studies were conducted on a FEI Tecnai F20
TEM/STEM equipped with both XEDS and electron energy loss spectroscopy (EELS), which was operated at 200 kV. In addition, high-resolution TEM images were obtained using an FEI Titan 80/300
field emission TEM operating at 300 kV. Fig \ref{Fig1} presents a  HAADF STEM  image of the heterostructure, demonstrating regularly shaped corners and equal layer thicknesses on both overgrown facets.

We observe stripes of high contrast along the diagonal of the intersection between perpendicular AlGaAs layers in the structure. Fig. \ref{Fig2} displays a HAADF STEM image of these features taken from the vicinity of region 1 in Fig. \ref{Fig1}, showing these high-contrast diagonal stripes in detail. The contrast in a HAADF image is dominated by atomic number effects, and higher average atomic number regions appear brighter due to their increased scattering cross-section \cite{HAADF}. In addition, in Fig. \ref{Fig2} we observe a secondary sub-structure of weaker but clearly resolved alternating bright and dark stripes in the growth direction of the individual AlGaAs layers. This indicates periodic variations of the Al content as the tilted facets rotate under the asymmetrically placed Al and Ga beams \cite{one_two}.
The material composition of the regions of interest was investigated with XEDS in Fig. \ref{Fig2} (inset) at three different locations: GaAs, AlGaAs and one of the diagonal stripes. Here the gray lines are the actual experimental data while the black lines are Gaussian peak fits to determine the local compositions. All spectra were normalized to the As L$_\alpha$ line, since we expect an equal As concentration in the three measured regions. The Al-content is estimated by a simple algebraic scaling of intensity ratios \cite{XEDS_ratio} from the stripe spectrum to the AlGaAs spectrum at a known average Al-content of x = 0.301. We obtain two independent estimates: $x^A = 0.78$ from the Al K$_{\alpha}$ intensities and $x^G = 0.51$ from the Ga L$_{\alpha}$ intensities. Using the mean value, we arrive at $\bar{x} = 0.65 \pm 0.1$ in the diagonal stripes, a significant increase compared to the neighboring $\mathrm{Al_{0.3}Ga_{0.7}As}$. The measurement error of the XEDS is $\pm\,10\,\%$, estimated from the Ga L$_\alpha$ intensities observed in Fig. \ref{Fig2} at known Ga concentrations in pure GaAs and AlGaAs.

\begin{figure}
\includegraphics{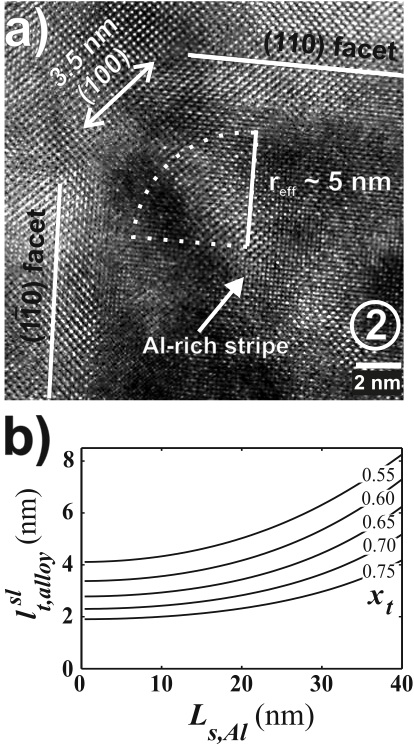}
\caption{Panel a shows a high-resolution bright-field TEM image of the the interface between layers 186/187 (AlGaAs/AlAs) in sample region 2 marked in Fig. \ref{Fig1}, at a total grown thickness of 1.2 $\mu$m. One of the Al-rich diagonal stripes investigated in Fig. \ref{Fig2} appears as a bright stripe of circa 2 nm width in this bright-field image. The interface to the AlAs layer 187 shows a sharp profile with a 3.5 nm wide (100) top facet. Panel b shows the self-limiting width calculated from Eq. \ref{lsl_alloy} for an ${\rm Al_{0.3}Ga_{0.7}As}$ alloy profile, as a function of the Al diffusion length $L_{s,Al}$ for various Al concentrations $x_t$ in the diagonal stripes.}
\label{Fig3}
\end{figure}

A high-resolution bright-field TEM image taken at the AlGaAs/AlAs interface between layers 186/187 (region 2 in Fig. \ref{Fig1}), after 1.2 $\mu$m of growth, reveals a sharp corner profile with evidence of a flat top of 3.5 nm width, which appears to be parallel to the $(100)$ facet (Fig. \ref{Fig3}a). Initial curvature of the overgrown corner, due to oxidation of the ex-situ cleaved substrate \cite{footnote1}, heals to this extremely sharp, faceted profile. The effective radius of curvature $r_\mathrm{eff}$ at the corner is a useful parameter for estimating quantum confinement, defined here as the radius of the smallest circle tangent to the $(110)$ and $(1\bar{1}0)$ side facets and the flat top region. We obtain $r_\mathrm{eff}\sim 5$ nm. The circle in Fig. \ref{Fig3}a is shifted away from the interface to give a clear view of the corner profile.

Both a faceted corner profile and the observed adatom segregation in AlGaAs layers are predicted in an analytic model by Biasiol {\it et al.} \cite{6_8,7}, explaining self-ordering of nanostructures grown on nonplanar surfaces. If the growth rates on the side facets $(r_s)$ and the top facet $(r_t)$ of a ridge profile show an anisotropy such that $\Delta r = r_s - r_t <0$, the faster growth on the side facets leads to a narrowing of the top facet. This effect is compensated by an increasing capillarity flux of adatoms away from the top facet, caused by an increase of the chemical potential at the narrowing top facet. The ridge structure develops a self-limiting profile of width $l_t^{sl}$, which for binary compounds like GaAs can be expressed by \cite{6_8,7}:
\begin{equation}
l_{t}^{sl}=\biggl(\frac{2\Omega_0L_{s}^2\gamma}{k_BT}\frac{r_{s}}{-\Delta r_{}}\biggr)^{1/3}
\label{lsl}
\end{equation}
Here,  $\Omega_0$ is the atomic volume of 11.8 ${\rm cm^3/mol}$, and $L_{s}$ is the adatom diffusion length on the sidewall.  The surface free energy term $\gamma = 2(\gamma_s\csc{\theta} - \gamma_t\cos{\theta})$ depends on the surface free energies of the top facet and the sidewall, $\gamma_t$ and $\gamma_s$ , and on the angle $\theta$ between the facets.
The scenario for ternary alloy layers, like the $\rm AlGaAs$ layer in Fig \ref{Fig2}, is more complicated: the different adatom diffusion lengths lead to different capillarity fluxes for the two adatom species, causing a region rich in the slow-diffusing species (Al) at the top facet. After Biasiol {\it et al.} \cite{7} the self-limiting width $l_{t,alloy}^{sl}$ is the solution to the equation 
\begin{equation}
\frac{a}{(l^{sl}_{t,alloy})^3}+\frac{b}{(l^{sl}_{t,alloy})^2}=\Delta r(x)
\label{lsl_alloy}
\end{equation}
with
\begin{eqnarray*}
a  &=&  x\Delta r_{Al}(l_{t,Al}^{sl})^3+(1-x)\Delta r_{Ga}(l_{t,Ga}^{sl})^3,\\
b & = & 2\big\{x(L_{s,Al})^2\ln(x/x_t) \\
&&+(1-x)(L_{s,Ga})^2\ln\left[(1-x)/(1-x_t)\right] \big\},\\
\Delta r(x) &=& x\Delta r_{Al} + (1-x)\Delta r_{Ga}
\end{eqnarray*}
where the subscripts Ga and As refer to the parameters for pure GaAs and AlAs growth, respectively. The logarithmic terms account for the entropy of mixing, with $x$ the Al content of the AlGaAs layer, and $x_t$ the local Al content at the top facet.

To compare our experimental findings to the predictions of this analytical growth model, we calculate the self-limiting width expected for an ${\rm Al_{0.3}Ga_{0.7}As}$ alloy layer from Eqs. \ref{lsl} and \ref{lsl_alloy}. From Ref. \cite{9} we estimate the surface free energies $\gamma_s\approx\gamma_t\approx45$ meV/\AA$^2$ under As$_4$ rich growth conditions, and obtain $\gamma = \sqrt{2}\gamma_s$ for the surface free energy term at $\theta = 45^{\circ}$. Experimental values for the  Ga diffusion length on the  (110) facet are found in Refs. \cite{10,11}. Extrapolating the temperature dependence measured by L\'{o}pez {\it et al.} \cite{10} at an As$_4$ pressure of $1.2\times 10^{-3}$ Pa, we obtain $L_s\approx40$ nm at our growth temperature of 460 $^{\circ}$C. 
The growth rate anisotropy $\Delta r_{Ga}$ is as well estimated from experimental results in Ref. \cite{10}, showing a reduced growth rate on (110) facets. We find a ratio of $1/0.23$ between (100) and (110) growth at our growth conditions. In addition, the growth rate at the edges of a (100) top facet is found to be locally enhanced by an additional adatom flux from the (110) side facets. We extrapolate an additional growth rate enhancement of 45 \% for a very narrow (100) top facet. For GaAs we therefore obtain a factor $\frac{r_{s,Ga}}{-\Delta r_{Ga}} \approx \frac{1}{1.45/0.23-1}$. Lacking experimental values, we assume the same growth rate anisotropy for AlAs growth, since the final result $l_{t,alloy}^{sl}$ is not very sensitive to this parameter \cite{footnote2}.
With the Al diffusion length $L_{s,Al}$ as the remaining free parameter, we calculate the self-limiting width $l_{t,alloy}^{sl}$ of an ${\rm Al_{0.3}Ga_{0.7}As}$ alloy layer for various values of the Al-content $x_t$ in the accumulation region. Fig. \ref{Fig3} shows $l_{t,alloy}^{sl}$ as a function of the Al diffusion length in the limits $0 < L_{s,Al} < L_{s,Ga} = 40$ nm. For $L_{s,Al} \ll L_{s,Ga}$, the measured top facet width of 3.5 nm would correspond to an Al-content of $x_t=0.6$, which is well within the error bars of our XEDS result $x_t = 0.65 \pm 0.1$. At larger values of the Al diffusion length, the measured width would still be within the limits of variation of the measured Al content $x_t$. This shows that the growth model in Refs. \cite{6_8,7} provides reasonable estimates of the Al segregation and the width of the self-limiting profiles created by MBE corner overgrowth.

The measured minimal curvature of the corner-overgrown layers also answers important questions about the electronic structure of bent quantum Hall junctions \cite{4}. For example, a one-dimensional accumulation wire has been predicted to exist at $B = 0$ in the structures of Ref. \cite{4}, if the diameter of curvature $2r$ is smaller than half the Fermi wavelength $\lambda_F $. For typical sheet electron densities between $1.0\times10^{11}\,{\rm cm}^{-2}$ and $1.5\times10^{11}\,{\rm cm}^{-2}$ $\lambda_F$ is between 80 nm and 65 nm, so from our measurements the condition $2r<\lambda_F/2$ is satisfied, and a one-dimensional wire with a single occupied subband should exist at the corner. To calculate the electronic dispersions in the presence of a $B$-field, the comparison between the magnetic length $l_B$ and the triangular confinement width $W$ becomes the relevant.  The wavefunction full-width at half-maximum $W = 18$ nm is estimated from Hartree calculations of the triangular confinement for the above densities. For small magnetic fields such that $W/2 < l_B$ ($B < 8$ T), the $B=0$ Hartree potential can be safely used in place of the finite $B$ Hartree potential, drastically simplifying the dispersion calculation as in Ref. \cite{4}.  At larger magnetic fields such that $W/2 > l_B$ ($B > 8$ T), dispersion calculations should include the $B$-field in the Hartree iteration.  Finally, at extreme fields the radius of curvature of the corner $r_\mathrm{eff} = 5$ nm becomes important once $r > l_B$ ($B > 26$ T).  At these high fields, the sharp corner approximation for the external potential should be replaced with a real potential with finite corner curvature. The corner-overgrown profiles measured here are sharp enough that one is able to achieve fractional filling factor $\nu = 1/3$ for typical densities before reaching this limit.

In conclusion, the TEM/STEM/XEDS results demonstrate that the corner overgrowth technique yields extremely sharp self-limited corner profiles, justifying the assumption of a sharp corner potential in bent quantum Hall junctions \cite{4}. Compared to the lower resolution TEM images of self-ordered diagonal stripes shown in Ref. \cite{7}, the stripes observed in our corner-overgrown structure are a factor of 5 thinner. Both the width of the self-limited profile and the Al adatom segregation observed in XEDS measurements on the AlGaAs layers can be estimated from a quantitative growth model \cite{6_8,7} with reasonable assumptions. Due to the self-limitation it is possible to grow sharp corner profiles that are translated over microns of growth, where the minimum width is determined by the growth conditions. This shows that the combination of MBE and corner overgrowth allows the precise control of one-dimensional quantum confinement, and justifies the assumptions of previous work of a sharp corner with a one-dimensional accumulation wire \cite{3,4}.  

\begin{acknowledgements}
\noindent This work was supported by the Deutsche Forschungsgemeinschaft by DFG GR 2618/1-1. In addition, some of this work was accomplished at the Electron Microscopy Center of Argonne National Laboratory, a U.S. Department of Energy Office of Science Laboratory operated under Contract No. DE-AC02-06CH11357 by UChicago Argonne, LLC.
\end{acknowledgements}

\noindent eMail: $^{\star}$lucia.steinke@wsi.tum.de,\\
\indent \indent \indent $^{\dagger}$m-grayson@northwestern.edu


\begin{thebibliography}{00}
\bibitem{footnote2} For example, at intermediate values $x_t = 0.65$ and $L_{s,Al} = 10$ nm, $l_{t,alloy}^{sl}$ varies only from 2.6 nm to 3.0 nm as the AlAs growth rate anisotropy is varied between -200 and -0.02.
\end{thebibliography}
\end{document}